\documentclass[a4paper,11pt]{article}
\pdfoutput=1

\usepackage{jheppub}
\usepackage[T1]{fontenc}
\usepackage{float}
\usepackage{amsmath,amssymb,graphics,epsfig,subfigure}
\usepackage{color}
\usepackage{hyperref}
\usepackage{epstopdf}
\usepackage{graphicx}

\newcommand \beq{\begin{equation}}
\newcommand \eeq{\end{equation}}
\newcommand \beqn{\begin{eqnarray}}
\newcommand \eeqn{\end{eqnarray}}
\newcommand \bseq{\begin{subequation}}
\newcommand \eseq {\end{subequations}}

\title{\boldmath Localization of Spinor Fields in Higher-Dimensional Braneworlds}

\author[a,b]{Jun-Jie Wan,}
\author[a,b,c,1]{and Yu-Xiao Liu \note{Corresponding author.}}

\affiliation[a]{Institute of Theoretical Physics $\&$ Research Center of Gravitation, Lanzhou University,\\Lanzhou 730000, P. R. China}
\affiliation[b]{Lanzhou Center for Theoretical Physics $\&$ Key
Laboratory of Theoretical Physics of Gansu Province, \\Lanzhou 730000, P. R. China}
\affiliation[c]{Key Laboratory for Magnetism and Magnetic Materials of the MOE, Lanzhou University,\\Lanzhou 730000, P. R. China}

\emailAdd{wanjj19@lzu.edu.cn}
\emailAdd{liuyx@lzu.edu.cn}

\abstract{
This study investigates the localization of spinor fields in braneworld models in six or higher dimensions.  {We study the reduction from a Dirac spinor in $2n+2$ dimensional spacetimes to spinors in $2n$ dimensions. The high-dimensional Dirac can be reduced to two Weyl spinors or Dirac spinors.} In conformally flat extra-dimensional spacetime, fermions cannot be localized through minimal coupling with gravity. To achieve the localization of spinor fields, we introduce a tensor coupling term given by $\bar{\Psi} \Gamma^{M}\Gamma^{N}\Gamma^{P}\cdots T_{MNP\cdots} \Psi$, which ensures $SO(n, 1)$ symmetry. For a tensor $T_{MNP\cdots}$ of odd order, the left and right chiralities of high-dimensional spinors are decoupled. We find that a special form of tensor coupling $\bar{\Psi}\Gamma^{M} \partial_M {F(\phi, R, R^{\mu\nu}R_{\mu\nu},\cdots)} \Psi$ may facilitate the localization of the spinor field when $F(\phi) = \phi^n$.
}

\begin{document}
\maketitle
\flushbottom

\section{Introduction}
If the existence of extra dimensions is assumed, how can a concise, self-consistent, and unified theory consistent with the current theoretical framework be constructed?

The idea of extra dimensions originated from attempts to unify different forces in nature. For example, in the Kaluza-Klein (KK) theory \cite{Kaluza:1921tu, Klein:1926tv}, the $4$-dimensional gravity and electromagnetism are described by a $5$-dimensional metric. This led to the idea of explaining fundamental interactions in terms of spacetime geometry. However, when the extra dimension is compactified into a loop, the reduction of the $5$-dimensional Fermion field does not result in a chiral theory in $4$ dimensions, which conflicts with the experimental observation that only left-handed neutrinos exist. Therefore, it is necessary to modify the KK theory in order to produce chiral fermions. The most common solution is to use orbifold compactification \cite{Cheng:2010pt}. For models with infinitely large extra dimensions or thick branes, an appropriate localization mechanism should also be used to obtain a $4$-dimensional chiral theory.

In theories of extra dimensions, it is possible that extra dimensions are non-compact and infinitely large, e.g., there is an infinitely large extra dimension in the 5-dimensional Randall-Sundrum (RS) braneworld model \cite{Randall:1999vf}. However, in this case, to ensure consistency with 4-dimensional observations, a localization mechanism must exist to confine matter fields within a specific spatial range, namely the brane in which we live. For such a model to be realistic, this mechanism could take forms such as a domain wall \cite{Rubakov:1983bb,Akama:1982jy}.

It is known that it is not possible to simultaneously localize free scalar and vector fields in a $5$-dimensional RS-like braneworld model. The most natural approach is the minimal coupling between matter field and gravity by introducing additional compact dimension to allow for the localization of the $U(1)$ gauge field \cite{Freitas:2018iil, Wan:2020smy}. This prompts us to consider $6$ or higher-dimensional spacetime. However, the localization of fermions in braneworld models with spacetime dimension greater than $5$ remains an area requiring systematic investigation.

In Ref.~\cite{Budinich:2001nh}, Budinich studied the relationship between high-dimensional pure spinor and $4$-dimensional spinor using Clifford algebra in flat spacetime. This work provides some insight into how fermions can be described in higher-dimensional spacetime. However, the spacetime with a braneworld is not flat. This means that, in addition to studying the algebraic structure of spinors, we also need to consider the generalized Dirac equation in higher-dimensional curved spacetime in order to construct a consistent theory of fermion dynamics.

 {In order to obtain $4$-dimensional free massless fermions, the KK zero mode must be localized on the brane. Reference \cite{Randjbar-Daemi:2000lem} proved that a $5$-dimensional massless Dirac fermion is generally non-normalizable with the use of minimal coupling of gravity and gauge fields under quite general assumptions about the geometry and topology of the internal manifold. The localization mechanism has been widely studied in the context of $5$-dimensional braneworld models \cite{Ringeval:2001cq, Melfo:2006hh, Flachi:2009uq, Chumbes:2010xg, Castillo-Felisola:2012fpz, Liu:2013kxz, Guo:2014nja, Li:2017dkw, MoazzenSorkhi:2018lvb}. }

 {To ensure the localization of the spinor zero mode, in a $5$-dimensional braneworld model,  the introduction of the Yukawa coupling allows for the localization of the fermion field on the brane.} Matter, gauge, and Higgs fields are allowed to propagate in the bulk, and the Standard Model fields and their interactions can be reproduced by the corresponding zero KK modes \cite{Dubovsky:2000am, Smolyakov:2015zsa}.  By including a Yukawa-type coupling to a scalar field of a domain-wall type, we can ensure the chirality as well as the localization of the fermions. However, these mechanisms do not work well in higher-dimensional braneworld models. In higher dimensions, if one carefully considers each component of a higher-dimensional spinor field, the localization of fermions presents a general difficulty. The left and right chiral massless KK modes are decoupled, but the massive KK modes are not. The coupling between the higher-dimensional left and right chiralities cannot be avoided as long as a mass term or a Yukawa coupling term is present; the interaction between $4$-dimensional fermions results from this coupling. This means that a mass term for higher-dimensional fermions prevents one from obtaining a $4$-dimensional free fermion.  {However, we hope that a reasonable effective theory should include a description of the free field, that is, a description of the propagation process of free particles.}

The aim of this paper is to provide general discussions applicable to higher-dimensional spacetime, rather than constructing a specific model.  {We hope to construct a localization mechanism that meets the following physical requirements:
\emph{\begin{itemize}
  \item (a) The high-dimensional spinor field and its interactions satisfy the Lorentz invariance.
  \item (b) Localization of the spinor zero mode, resulting in a low-dimensional effective theory.
  \item (c) In the effective theory, a description of an effective low-dimensional free field is required.
\end{itemize}}}
We will study whether fermions can be naturally localized and obtain $4$-dimensional free massless fermions by introducing a coupling term into the higher-dimensional fundamental theory and decouple the left and right chiralities of higher-dimensional spinors. We require that this decoupling should not break the $SO(n, 1)$ symmetry. We hope to obtain a $4$-dimensional chiral theory  through a reasonable localization mechanism. Therefore, we take the background of different topologies in the $6$-dimensional braneworld as an example to study the distinction in fermion chirality.

This paper is organized as follows. In section \ref{Fermion_localization}, we introduce a general localization mechanism, study the kinetic equations of high-dimensional spinors, and present the results. In section \ref{Examples}, we take $6$-dimensional spacetime as an example. Under some specific coupling forms, we study the localization conditions of the fermion zero modes. Differences in the chirality of fermions under different spacetime topological backgrounds are discussed. The last section \ref{Conclusions_Outlook} is devoted to conclusions and outlook.  {In the appendix, we will provide the mathematical foundations of Clifford algebra and high-dimensional spinors that are needed for this paper, for reference.}

\section{Fermion localization in high-dimensional spacetime}\label{Fermion_localization}

To achieve a localization mechanism, one might consider the coupling between the spinor and the background dynamic field. Such a coupling can greatly facilitate localization, particularly in highly symmetric spacetimes, where the interaction term ensures the material field distribution is consistent with the symmetry of the brane.

To ensure the invariance of the Lagrangian under a Lorentz transformation, a spinor and its adjoint should be combined to form the Lorentz scalar $\bar{\Psi} \Psi$ or the Lorentz vector $\bar{\Psi} \Gamma^{A} \Psi$, as well as other combinations. First, it is natural to consider the following action for the spinor in a $(2n+2)$-dimensional spacetime with a Yukawa-like coupling \cite{Fu:2011pu, Castillo-Felisola:2012fpz, Barbosa-Cendejas:2015qaa, Guo:2014nja, Flachi:2009uq, Salvio:2007qx, Chumbes:2010xg, Slatyer:2006un, Kodama:2008xm, Ringeval:2001cq, Melfo:2006hh}. We introduce
\begin{eqnarray}\label{Coupled_to_scalar}
{S_{1}} = \int {d}^{2n+2} {x}~ \sqrt{-{g}} \left[\bar{\Psi} \Gamma^{M} {D}_{{M}} \Psi
+ \eta F(\phi, R, R^{\mu\nu}R_{\mu\nu},\cdots) \bar{\Psi}
\Psi \right],
\end{eqnarray}
where $F(\phi, R, R^{\mu\nu}R_{\mu\nu},\cdots)$ represents a scalar function dependent on the scalar field $\phi$, the curvature scalar $R$, and other geometric scalars.

 {The introduction of the mass term or Yukawa coupling couples the left- and right-handed chiralities of the fermion. The left- and right-handed chiralities of the fermion transform as independent entities under the Lorentz transformation~\eqref{Lorentz_transformation}. This prohibits decoupling the left- and right-handed chiralities through coordinate selection. In order to obtain a $4$-dimensional effective free field, we aim to find the $4$ components that have independent equations of motion from other components of the high-dimensional spinor. The Yukawa term appears to introduce a contradiction between high-dimensional Lorentz symmetry and $4$-dimensional free field theory.}

Given a coupling between left and right chiral spinors in a $(2n+2)$-dimensional fundamental theory, a $2n$-dimensional free field cannot be derived through action reduction from either the left or right chiral spinors. One potential solution involves inserting an odd number of Gamma matrices between $\bar{\Psi}$ and $\Psi$, as exemplified by
\begin{eqnarray}\label{PsiGammaMatricesPsi}
  \bar{\Psi} \Gamma_{A_1}\Gamma_{A_2}\cdots\Gamma_{A_{2k+1}}\Psi,
\end{eqnarray}
 {where the indices are fixed as $A_1,A_2,\cdots,A_{2k+1}$, such as $\bar{\Psi} \Gamma_{2n}\Psi$ or $\bar{\Psi} \Gamma_{0}\Gamma_{1}\Gamma_{2}\Psi$. The coupling forms \eqref{PsiGammaMatricesPsi} do not possess Lorentz invariance.
A spinor and its adjoint must combine to form a Lorentz scalar. Next, we introduce the form of the interaction $$\bar{\Psi} \Gamma^{M}\Gamma^{N}\Gamma^{P}\cdots T_{MNP\cdots} \Psi$$ to resolve this contradiction.}

%Similarly, as in the 5-dimensional braneworld model, we can distinguish lower-dimensional left and right chiral spinors using a Yukawa-like coupling.
%\textcolor[rgb]{0.00,0.00,1.00}{While it is evident that terms such as $\bar{\Psi} \Gamma_{2n}\Psi$ are not Lorentz-invariant, the following conditions cannot be met simultaneously for the above constructions:
%\begin{itemize}
%  \item (a)
%  \item (b) Localization of the spinor zero mode, resulting in a low-dimensional effective theory. This requires that the bulk spinor should couple with other fields.
%  \item (c) An effective low-dimensional free field theory. This implies that left and right chiral fermions in higher dimensions decouple from each other. (This condition is not satisfied for the coupling term $F(\phi, R, R^{\mu\nu}R_{\mu\nu},\cdots) \bar{\Psi} \Psi$.)
%\end{itemize}}

A possible solution involves imposing the type of coupling $\bar{\Psi} \Gamma^{M} \xi_M \Psi$, where $\xi_M$ is a vector. We can also extend the coupling to an odd-order tensor field, for example, $\bar{\Psi} \Gamma^{M}\Gamma^{N}\Gamma^{P} T_{MNP} \Psi$. Based on the preceding analysis, we focus on coupling with a vector field $\xi_M$ as opposed to a scalar function $F$. The action takes the form
\begin{eqnarray}
{S_{2}}=\int {d}^{2n+2} {x}~\sqrt{-{g}}\left[\bar{\Psi} \Gamma^{M} {D}_{{M}} \Psi
\right.+ \left. \varepsilon  \bar{\Psi} \Gamma^{M} \xi_M \Psi \right].
\end{eqnarray}
The Dirac equation is given by
\begin{align}
\left[ \partial_{M}+\Omega_{M}+ \varepsilon  \xi_M \right] \Gamma^{M} \Psi\left(x^{N}\right)=0.
\end{align}
Consider the 'Weyl' representation of the Gamma matrices and note that $\Gamma_{A} \Gamma_{B}$ are block diagonal matrices. We have
\begin{eqnarray}
\Omega_{M}=\frac{1}{4} \Omega_{M}^{A B} \Gamma_{A} \Gamma_{B}=\left(\begin{array}{cc}
\omega_{1M} & 0 \\
0 & \omega_{2M}
\end{array}\right),
\end{eqnarray}
where the spin connection $\Omega_{M}^{A B}$ is defined in Eq.~(\ref{spinconnection}). From Eq.~\eqref{Diraceq}, we find that the left and right chirality fermions satisfy independent equations of motion with the coupling term $\bar{\Psi} \Gamma^{M} \xi_M \Psi$.

 {
First, we consider the general form of the diagonal metric to obtain mathematically general results. For such a metric, the line element \( ds^2 \) is expressed as:}
\begin{align}\label{line elements}
ds^2= -a_0^2(x^{N}) dx_0^2+\sum_{k=1}^{2n+1} a_k^2(x^{N}) dx_k^2,
\end{align}
where \( a_0(x^{N}) \) and \( a_k(x^{N}) \) are the warp factors for all spacetime coordinates. The spin connection \( \Omega_{j} \) is given by
\begin{align}
\Omega_{j}=\frac{1}{2} \Gamma_j \sum_{i=0, i\neq j}^{2n+1}  \frac{\Gamma^i \partial_i a_j(x^{N})}{a_i(x^{N})}.
\end{align}
Furthermore, we have
\begin{align}
\Gamma^{M}\Omega_{M}=\frac{1}{2}\sum_{j=0}^{2n+1}\sum_{i=0,i\neq j}^{2n+1}\frac{\Gamma^i\partial_{i}a_j(x^{N})}{a_i(x^{N}) a_j(x^{N})}.
\end{align}
A vector field $G_M$ can be defined, its components $G_i$ are given by
\begin{align}
G_i=\frac{1}{2}  \sum_{j=0,i\neq j}^{2n+1}\frac{\partial_{i}a_j(x^{N})}{a_j(x^{N})}.
\end{align}
Although $G_M$ is not the spin connection and its components $G_i$ are functions depending on the choice of coordinates, $G_M$ serves as an alternative to the spin connection in the Dirac equation within the coordinate framework established by Eq.~\eqref{line elements}. Therefore, we have
\begin{align}
\Gamma^{M}\Omega_{M} = \Gamma^{A} E_{A}^{M}  G_{M};
\end{align}
subsequently, the Dirac equation takes the form
\begin{equation}\label{Diraceq310}
	\left[ \Gamma^{{A}}(E_{A}^{M} \partial_{M}+ E_{A}^{M} G_M)+  \varepsilon  \Gamma^M \xi_M  \right] \Psi\left(\xi_{N}\right)=0.
\end{equation}

As discussed in the previous section, $\bar{\Psi} \Gamma^{M} \xi_M \Psi$ does not couple the left and right chiralities. This decoupling enables the existence of two independent $4$-dimensional Dirac free fields in the $4$-dimensional effective theory. Utilizing the Weyl representation, we can decompose Eq.~\eqref{Diraceq310} into two independent components; these correspond to the left chiral spinor $\Psi_L$ and the right chiral spinor $\Psi_R$, as given by
\begin{align}
\hat{D}^{(2\iota)}_L = P_R \Gamma^{{A}} E_{A}^{M} (\partial_{M}+\varepsilon\xi_M  + G_M)  , \\
\hat{D}^{(2\iota)}_R = P_L \Gamma^{{A}} E_{A}^{M} (\partial_{M}+\varepsilon\xi_M  + G_M)  ,
\end{align}
where $\hat{D}_L$ and $\hat{D}_R$ are operators that act on $\Psi_{L}$ and $\Psi_{R}$, respectively, as indicated in Eq.~\eqref{Diraceq310}. Employing these operators, we rewrite the equation as
\begin{align}
\hat{D}^{(2\iota)}_L \Psi_L^{(2 \iota)}=\hat{D}^{(2\iota)}_L \left(\begin{array}{l}
\Psi_{1}^{(\iota)} \\
0
\end{array}\right)=0, \\
\hat{D}^{(2\iota)}_R \Psi_R^{(2 \iota)}=\hat{D}^{(2\iota)}_R \left(\begin{array}{l}
0 \\
\Psi_{2}^{(\iota)}
\end{array}\right)=0.
\end{align}
It is noteworthy that the matrix operators $P_R \Gamma^{{A}}$ and $P_L \Gamma^{{A}}$ are degenerate. As a result, the dimension can transition from $2 \iota$ to $\iota$. We refer to the degenerate operators as $\hat{D}^{(\iota)}_L$ and $\hat{D}^{(\iota)}_R$, which operate solely on the two independent $2^n$-component spinors ${\Psi}^{(\iota)}_1$ and ${\Psi}^{(\iota)}_2$, respectively.

The action corresponding to this decomposition can be further split into its right and left chiral components as follows:
\begin{align}
{S}=S_L+S_R,
\end{align}
where \(S_L\) and \(S_R\) are defined by
\begin{align}
{S_L} = \int {d}^{2n} {x} \sqrt{-{g}} \left[\bar{\Psi}_1^{(\iota)} {\hat{D}}^{(\iota)}_{{L}}{\Psi}^{(\iota)}_1 \right],
\end{align}
and
\begin{align}
{S_R} = \int {d}^{2n} {x} \sqrt{-{g}} \left[\bar{\Psi}_2^{(\iota)} {\hat{D}}^{(\iota)}_{{R}}{\Psi}^{(\iota)}_2 \right].
\end{align}

We start by defining the operators
\begin{align}
\tilde{D}^{(2\iota)}_L = P_R \Gamma^{{A}}E_{A}^{M} \partial_{M} , \\
\tilde{D}^{(2\iota)}_R = P_L \Gamma^{{A}}E_{A}^{M} \partial_{M} .
\end{align}
For the diagonal metric, and further representing the frame using warp factors, we find
\begin{align}
\tilde{D}^{(2\iota)}_L = P_R \Gamma^{{i}} \tilde{\partial}_{i} , \\
\tilde{D}^{(2\iota)}_R = P_L \Gamma^{{i}} \tilde{\partial}_{i} ,
\end{align}
where \(\tilde{\partial}_{i} \equiv a_{i}^{-1}  \partial_{i}\) (no sum over \(i\)), and we define \(\tilde{\xi}_i \equiv \varepsilon a_i^{-1} \xi_i\) (no sum over \(i\)) and \(\tilde{G}_i = {a_i}^{-1} {G}_i\).
This transformation of the field allows us to describe the interaction of the field \(\xi_M\) in curved spacetime in the same manner as in flat spacetime.
It should be noted that these operators, \(\tilde{D}^{(2\iota)}_L\) and \(\tilde{D}^{(2\iota)}_R\), are applied only to what we will refer to as the "half-component" of the spinors. Thus, the non-zero elements of these operators can be denoted as \(\tilde{D}^{(\iota)}_L\) and \(\tilde{D}^{(\iota)}_R\).

To illustrate the reduction from $6$ to $4$ dimensions, we examine the following action terms. The action \(S_L\) is given by
\begin{equation}
\begin{aligned}
S_{L} &= \int {d}^{6} {x} \sqrt{-{g}}
\left[\bar{\Psi}_1^{(\iota)} \tilde{D}^{(\iota)}_{{L}}{\Psi}^{(\iota)}_1 +
\bar {\Psi}_1^{(\iota)} (\tilde{\xi}_a + \tilde{G}_a) \gamma^a \Psi_1^{(\iota)} \right. \\
& \left. + \bar {\Psi}_1^{(\iota)} (\tilde{\xi}_5 + \tilde{G}_5) \gamma^5 \Psi_1^{(\iota)}
+ i \bar {\Psi}_1^{(\iota)} (\tilde{\xi}_6 + \tilde{G}_6) \Psi_1^{(\iota)}
\right],
\end{aligned}
\end{equation}
while the action \(S_R\) is
\begin{equation}
\begin{aligned}
S_{R} &= \int {d}^{6} {x} \sqrt{-{g}}
\left[\bar{\Psi}_2^{(\iota)} \tilde{D}^{(\iota)}_{{R}}{\Psi}^{(\iota)}_2 +
\bar {\Psi}_2^{(\iota)} (\tilde{\xi}_a + \tilde{G}_a) \gamma^a \Psi_2^{(\iota)} \right. \\
& \left. + \bar {\Psi}_2^{(\iota)} (\tilde{\xi}_5 + \tilde{G}_5) \gamma^5 \Psi_2^{(\iota)}
- i \bar {\Psi}_2^{(\iota)} (\tilde{\xi}_6 + \tilde{G}_6) \Psi_2^{(\iota)}
\right],
\end{aligned}
\end{equation}
where \(\gamma^a\) are generators representing the subalgebra of the Clifford algebra. This action is partitioned into \(S_L\) and \(S_R\), which correspond to \({\Psi}^{(\iota)}_1\)  and \({\Psi}^{(\iota)}_2\), respectively.

Though the actions \(S_L\) and \(S_R\) appear as \(4\)-dimensional actions, it should be noted that \({\Psi}_1\) and \({\Psi}_2\) remain as spinor fields in the bulk. For a complete reduction to \(4\)-dimensional effective actions, the KK decomposition is also required.

 {The diagonal metric Eq.~\eqref{line elements} is a relatively common situation. However, in the context of the brane-world scenario, the space of extra dimensions and the four-dimensional space can be decomposed independently. This decomposition facilitates the formulation of a 4-dimensional effective theory. Notably, the four-dimensional space is conformally flat. By assuming a flat brane, our focus is directed towards the localization of the matter field on the brane, avoiding the complexities of the brane's internal structure. In the context of braneworld models within high-dimensional spacetimes, the additional dimensions are typically characterized by a pronounced symmetry. For scenarios involving two extra dimensions, this manifests as a specific orientation in the additional dimensional space, denoted as \(x^5\).
Considering the metric:
\begin{equation}\label{line element_x5}
    d s^2 = a_4(x^5) \eta_{\mu\nu} d x^\mu d x^\nu + a_5(x^5) d x^5 d x^5 + a_6(x^5) d x^6 d x^6,
\end{equation}
%where \( y \) represents the extra-dimensional coordinates \( (x^4, x^5) \).
%\begin{itemize}
%  \item \( a_4(y) \equiv a_0(y) = a_1(y) = a_2(y) = a_3(y) \),
%  \item \( \xi_4(y) \equiv \xi_0(y) = \xi_1(y) = \xi_2(y) = \xi_3(y) \).
%\end{itemize}
%\begin{align}
%a(x^5) &= a(x^5), \\
%b(x^5) &= b(x^5),
%\end{align}
%and
%\begin{align}
%F(x^5, x^6) = F(x^5).
%\end{align}
The integral of the Lagrangian can be decomposed into $4$ dimensional and extra dimensional
%\begin{equation}
%\begin{split}
%S_{L, R}=&\int{dx^5}dx^6\tilde{\alpha}_4(x^5,x^6) \int{d^4}x\,\,\bar{\psi}_{1,2}^{(\iota )}(x^a)\gamma ^a\partial _a\psi _{1,2}^{(\iota )}(x^a)
%\\
%&+\int{dx^5}dx^6\tilde{\alpha}_5(x^5,x^6) \frac{\partial _5\phi_{1,2}(x^5, x^6)}{\phi_{1,2}(x^5, x^6)}\int{d^4}x\,\,\bar{\psi}_{1,2}^{(\iota )}(x^a)\gamma ^5\psi _{1,2}^{(\iota )}(x^a)
%\\
%&\pm \int{dx^5}dx^6\tilde{\alpha}_6(x^5,x^6) \frac{\partial _6\phi _{1,2}(x^5, x^6)}{\phi _{1,2}(x^5, x^6)}\int{d^4}x\,\, \bar{\psi}_{1,2}^{(\iota )}(x^a) i\,\,\psi _{1,2}^{(\iota )}(x^a)
%\\
%&+\sum_{a=0}^3{\int{dx^5}dx^6 \tilde{\alpha}_4(x^5,x^6)  \left[\varepsilon \xi _4(x^5,x^6)+{G}_4(x^5,x^6)\right]  \int{d^4}x\,\,i\,\,\bar{\psi}_{1,2}^{(\iota )}(x^a)\gamma ^a\psi _{1,2}^{(\iota )}(x^a)}
%\\
%&+\int{dx^5}dx^6\tilde{\alpha}_5(x^5,x^6)  \left[ \varepsilon \xi _5(x^5,x^6)+ {G}_5(x^5,x^6)\right]  \,\,\int{d^4}x\,\,i\,\,\bar{\psi}_{1,2}^{(\iota )}(x^a)\gamma ^5\psi _{1,2}^{(\iota )}(x^a)
%\\
%&\pm \int{dx^5}dx^6\tilde{\alpha}_6(x^5,x^6)  \left[ \varepsilon \xi _6(x^5,x^6) + {G}_6(x^5,x^6)\right] \int{d^4}x\,\, i\,\,\bar{\psi}_{1,2}^{(\iota )}(x^a) i\,\,\psi _{1,2}^{(\iota )}(x^a),
%\end{split}
%\end{equation}
\begin{equation}
\begin{split}
S_{L, R}=& \int{d^4}x\,\,\bar{\psi}_{1,2}^{(\iota )}(x^a)\gamma ^a\partial _a\psi _{1,2}^{(\iota )}(x^a)
\\
&+\tilde{\alpha}_5 \frac{\partial _5\phi_{1,2}}{\phi_{1,2}}\int{d^4}x\,\,\bar{\psi}_{1,2}^{(\iota )}(x^a)\gamma ^5\psi _{1,2}^{(\iota )}(x^a)
\\
&\pm \tilde{\alpha}_6 \frac{\partial _6\phi _{1,2}}{\phi _{1,2}}\int{d^4}x\,\, \bar{\psi}_{1,2}^{(\iota )}(x^a) i\,\,\psi _{1,2}^{(\iota )}(x^a)
\\
&+\sum_{a=0}^3{\tilde{\alpha}_4  \left[\varepsilon \xi _4+{G}_4\right]  \int{d^4}x\,\,i\,\,\bar{\psi}_{1,2}^{(\iota )}(x^a)\gamma ^a\psi _{1,2}^{(\iota )}(x^a)}
\\
&+\tilde{\alpha}_5  \left[ \varepsilon \xi _5 + {G}_5\right]  \,\,\int{d^4}x\,\,i\,\,\bar{\psi}_{1,2}^{(\iota )}(x^a)\gamma ^5\psi _{1,2}^{(\iota )}(x^a)
\\
&\pm \tilde{\alpha}_6  \left[ \varepsilon \xi _6 + {G}_6\right] \int{d^4}x\,\, i\,\,\bar{\psi}_{1,2}^{(\iota )}(x^a) i\,\,\psi _{1,2}^{(\iota )}(x^a),
\end{split}
\end{equation}
where, we define
\begin{equation}
\tilde{\alpha}_{4,5,6} =\sqrt{-g}\left( \phi _{1}^{*}\phi _1\right) a_{4,5,6}^{-1}.
\end{equation}
Under the premise of metric \eqref{line element_x5}, we have
\begin{align}
G_4 &\equiv G_0 = G_1 = G_2 = G_3 = 0\\
G_5 &=  \frac{1}{2}  \sum_{\substack{j=1 \\ j\neq 5}}^{2n} {\partial_{5} \ln a_j}, \\
G_6 &= 0.
\end{align}
Corresponding to the kinetic energy term in $4$-dimensional effective theory, the localization condition is
\begin{equation}
\int{dx^5}dx^6\tilde{\alpha}_4 = 1.
\end{equation}
}
In the context of high-dimensional spacetime, we aim to derive $4$-dimensional free fields while preserving the underlying symmetries. A common approach is to couple bulk fermion and vector fields in the form \(\bar{\Psi} \Gamma^{M} \xi_M \Psi\). However, brane-world models often utilize a scalar field to characterize the dynamical background. One can construct a vector field from the derivative of such a scalar field, allowing us to use the covariant derivative of the scalar field or a scalar function based on geometric background quantities as the vector field coupling to the fermion. This setup provides a mechanism for localizing matter fields through derivative coupling, a topic that has been explored in literature \cite{Liu:2013kxz, Li:2017dkw}. It is worth noting that the couplings used in these localization mechanisms can be considered as special cases of the more general coupling scheme \(\bar{\Psi} \Gamma^{M}\Gamma^{N}\Gamma^{P}\cdots T_{MNP\cdots} \Psi\).

 {In the proposed model, the action takes the form}
\begin{equation}\label{action of vector localization mechanism}
S_{3} = \int {d}^{6} x \sqrt{-g} \left[ \bar{\Psi} \Gamma^{M} {D}_{M} \Psi + \varepsilon \bar{\Psi} \Gamma^{M} \partial_{M} F(\phi, R, R^{\mu\nu}R_{\mu\nu}, \ldots) \Psi \right],
\end{equation}
where the vector field \( \xi_M(x^5) \) is expressed as the derivative of a scalar function \( F(x^5) \) .
This scalar function \( F \) is a function of the scalar field \( \phi \), the Ricci scalar \( R \), and other curvature invariants like \( R^{\mu\nu}R_{\mu\nu} \), among other possible terms \footnote{ {The \(\xi^M\) in equation is actually a vector function constructed from the background dynamical fields and geometric scalars, rather than a vector constructed from the spinor field. It does not act as an independent dynamical field or degree of freedom. This vector field appears as a spacetime background field, and its dynamical terms are unrelated to the spinor field under consideration, and are only related to the dynamical description of the spacetime background. When considering the localization of fermions, we did not take into account the back-reaction of fermions on the spacetime background. As a specific example, the dynamics of the spacetime background are described by \eqref{Lagrangian of the background scalar} and \eqref{action}.}}. Under this assumption, we have $\xi _4=0$ and the action can be written as
\begin{eqnarray}
S_{L,R} &=&\int{dx^5}dx^6\tilde{\alpha}_4\,\,\int{d^4}x\,\,\bar{\psi}_{1,2}^{(\iota )}(x^a)\gamma ^a\partial _a\psi _{1,2}^{(\iota )}(x^a) \nonumber \\
&&+\int{dx^5}dx^6\tilde{\alpha}_5\,\,\left( \frac{\partial _5\phi _1}{\phi _1}\,+\varepsilon \partial _5F+ {G}_5 \right) \int{d^4}x\,\bar{\psi}_{1,2}^{(\iota )}(x^a)\gamma ^5\psi _{1,2}^{(\iota )}(x^a)
\nonumber \\
&& {\mp \int{dx^5}dx^6\tilde{\alpha}_6  \frac{\partial _6\varphi _1}{\varphi _1} \int{d^4}x\,\bar{\psi}_{1,2}^{(\iota )}(x^a)\psi _{1,2}^{(\iota )}(x^a)}.
 \label{4 dimensional effective action}
\end{eqnarray}
Note that the left- and right-handed parts of the high-dimensional spinor corresponds to independent Dirac spinors in the reduced effective action. The action not only includes zero-mass Dirac fermions but also a series of massive ones. The introduction of higher-dimensional spacetime provides us with an explanation for the generation mechanism of particle masses, which is distinct from the Higgs mechanism. This stands in contrast to the decomposition of a Dirac spinor into Weyl spinors, where the Weyl spinor is massless. In the effective four-dimensional theory, there are multiple massive particles, which are four-dimensional Dirac spinors, not Weyl spinors, even though we employ chiral representations in the study of higher-dimensional spinors.

In the coupling form $\bar{\Psi} \Gamma^{M} \xi_M \Psi$, both the left- and right-handed components of the higher-dimensional fermionic field satisfy independent equations of motion. Therefore, the covariant derivative can be defined with respect to either the left- or right-handed part of the higher-dimensional spinor. Furthermore, the equations of motion can be separated into $4$-dimensional and extra-dimensional parts. We can then separate the variables as follows:
\begin{align}
\hat{D}^{(2\iota)}_L &= \hat{D}^{(2\iota)}_{L brane} + \hat{D}^{(2\iota)}_{L extra}, \\
\hat{D}^{(2\iota)}_R &= \hat{D}^{(2\iota)}_{R brane} + \hat{D}^{(2\iota)}_{R extra},
\end{align}
where
\begin{align}
\hat{D}^{(2\iota)}_{L brane} &=  P_L \Gamma^{{A}} E_{A}^{M} (\partial_{M}+\varepsilon\xi_M  + G_M), ~ M=0,1,2,3, \\
\hat{D}^{(2\iota)}_{L extra} &=  P_L \Gamma^{{A}} E_{A}^{M} (\partial_{M}+\varepsilon\xi_M  + G_M), ~ M=5,6, \\
\hat{D}^{(2\iota)}_{R brane} &=  P_R \Gamma^{{A}} E_{A}^{M} (\partial_{M}+\varepsilon\xi_M  + G_M), ~ M=0,1,2,3, \\
\hat{D}^{(2\iota)}_{R extra} &=  P_R \Gamma^{{A}} E_{A}^{M} (\partial_{M}+\varepsilon\xi_M  + G_M), ~ M=5,6.
\end{align}
In light of the $4$-dimensional effective action~\eqref{4 dimensional effective action}, $\hat{D}_{\text{extra}}$ can be considered as a 'mass' operator. $\psi_{1}$ and $\psi_{2}$ serve to represent the $4$-dimensional part of higher-dimensional left- and right-hand particles, respectively. These correspond to a set of $4$-dimensional massless and massive fermions in the effective theory.

The $6$-dimensional Dirac equation can be expressed as
 {\begin{align}\label{Diraceq}
\left[ a_4^{-1}\Gamma^{\mu} \partial_{\mu} + a_5^{-1}\Gamma^{5} \left( \partial_{5} + G_5 + \varepsilon \partial_{5} F \right) + a_6^{-1}\Gamma^{6} \partial_{6} \right] \Psi\left( x^{N} \right) = 0,
\end{align}}
where \( G_5, G_6, \) and \( F \) are functions of \( x^5 \) and \( x^6 \).
Through the decomposition of \eqref{Chiral_decomposition_of_spinor} under the Weyl representation, the $6$-dimensional Dirac equation can be split into two components:
 {\begin{equation}
\begin{array}{l}
\left[ a_4^{-1}\gamma^{\mu}\partial_{\mu} + a_5^{-1}\gamma^{5} \left( \partial_{5} + G_5 + \varepsilon \partial_{5} F \right) + a_6^{-1}i \partial_{6} \right] \Psi_{1}^{(4)} = 0, \\
\left[ a_4^{-1}\gamma^{\mu}\partial_{\mu} + a_5^{-1}\gamma^{5} \left( \partial_{5} + G_5 + \varepsilon \partial_{5} F \right) - a_6^{-1}i \partial_{6} \right] \Psi_{2}^{(4)} = 0.
\end{array}
\end{equation}}
At this point, the $8$-component spinor decomposes into two independent $4$-component spinors. Each $4$-component spinor further decomposes into two $2$-component spinors by the following decomposition:
\begin{align}
\Psi_{1}^{(4)} = \left(\begin{array}{l}
\Psi_{11}^{(2)} \\
\Psi_{12}^{(2)}
\end{array}\right)
& = \sum_{m_1}\left(\begin{array}{l}
\psi_{11{m_1}}^{(2)}(x) \phi_{11{m_1}}^{(2)} e^{i l \Theta} \\
\psi_{12{m_1}}^{(2)}(x) \phi_{12{m_1}}^{(2)} e^{i l \Theta}
\end{array}\right), \nonumber \\
\Psi_{2}^{(4)} = \left(\begin{array}{l}
\Psi_{21}^{(2)} \\
\Psi_{22}^{(2)}
\end{array}\right)
& = \sum_{m_2}\left(\begin{array}{l}
\psi_{21{m_2}}^{(2)}(x) \phi_{21{m_2}}^{(2)} e^{i l \Theta} \\
\psi_{22{m_2}}^{(2)}(x) \phi_{22{m_2}}^{(2)} e^{i l \Theta}
\end{array}\right).
\end{align}

As analyzed in the previous section, the reduction of the action of the Dirac spinor field in higher-dimensional spacetime yields two independent $4$-dimensional Dirac spinors.
If we take the $4$-dimensional Weyl representation, then we have
\begin{align*}
    \gamma^{\mu} \partial_{\mu} \psi_L^{(4)} &=  m\psi_R^{(4)}, &
    \gamma^{\mu} \partial_{\mu} \psi_R^{(4)} &= m\psi_L^{(4)}, \\
    \gamma^5 \psi_L^{(4)} &=  \psi_L^{(4)}, &
    \gamma^5 \psi_R^{(4)} &= -\psi_R^{(4)},
\end{align*}
and
\begin{subequations}
\begin{align}
a_4^{-1} m_1\phi _{12}+ a_5^{-1} \left[ \partial _5+H_5 \right] \phi _{11}+i a_6^{-1} \partial _6 \phi _{11} &= 0, \\
a_4^{-1} m_1\phi _{11}- a_5^{-1} \left[ \partial _5+H_5 \right] \phi _{12}+i a_6^{-1} \partial _6 \phi _{12} &= 0, \\
a_4^{-1} m_2\phi _{22}+ a_5^{-1} \left[ \partial _5+H_5 \right] \phi _{21}-i a_6^{-1} \partial _6 \phi _{21} &= 0, \\
a_4^{-1} m_2\phi _{21}- a_5^{-1} \left[ \partial _5+H_5 \right] \phi _{22}-i a_6^{-1} \partial _6 \phi _{22} &= 0.
\end{align}
\end{subequations}
%\begin{subequations}
%\begin{align}
%a_4^{-1} m_1\phi _{12}+ a_5^{-1} \left[ \partial _5+H_5 \right] \phi _{11}+i a_6^{-1}\left[ \partial _6+H_6 \right] \phi _{11} &= 0, \\
%a_4^{-1} m_1\phi _{11}- a_5^{-1} \left[ \partial _5+H_5 \right] \phi _{12}+i a_6^{-1} \left[ \partial _6+H_6 \right] \phi _{12} &= 0, \\
%a_4^{-1} m_2\phi _{22}+ a_5^{-1} \left[ \partial _5+H_5 \right] \phi _{21}-i a_6^{-1} \left[ \partial _6+H_6 \right] \phi _{21} &= 0, \\
%a_4^{-1} m_2\phi _{21}- a_5^{-1} \left[ \partial _5+H_5 \right] \phi _{22}-i a_6^{-1} \left[ \partial _6+H_6 \right] \phi _{22} &= 0,
%\end{align}
%\end{subequations}
where we have defined
\begin{align}
H_5(x^5, x^6) &= G_5(x^5, x^6) + \varepsilon \partial _5F(x^5, x^6).
%H_6(x^5, x^6) &= G_6(x^5, x^6) + \varepsilon \partial _6F(x^5, x^6).
\end{align}
 {We can always do a coordinate transformation
\begin{align}
a_4^{-1} \tilde{\partial} _5 = a_5^{-1} \partial _5.
\end{align}
which gives $a_5=a_4$.}
Then, the equation can be written as
\begin{subequations}\label{h5}
\begin{align}
m_1\phi _{12}+ \left[ \partial _5+H_5 \right] \phi _{11}+i a_4 a_6^{-1} \partial _6 \phi _{11} &= 0, \\
m_1\phi _{11}- \left[ \partial _5+H_5 \right] \phi _{12}+i a_4 a_6^{-1} \partial _6 \phi _{12} &= 0, \\
m_2\phi _{22}+ \left[ \partial _5+H_5 \right] \phi _{21}-i a_4 a_6^{-1} \partial _6 \phi _{21} &= 0, \\
m_2\phi _{21}- \left[ \partial _5+H_5 \right] \phi _{22}-i a_4 a_6^{-1} \partial _6 \phi _{22} &= 0.
\end{align}
\end{subequations}
This includes the case where the bulk spacetime is conformally flat.
The effect of \(H_5\) can be equivalently described by a field transformation.
The field is transformed by the conformal factor as follows:
\begin{align}
\phi_{ij} = \Upsilon(x^5) \tilde{\phi}_{ij},
\end{align}
where the conformal transformation factor of the field is given by
\begin{align}
\Upsilon(x^5) = \exp\left(-\int H_5(x^5) dx^5\right) = \exp\left(-\int G_5(x^5) dx^5\right) \exp\left(-\varepsilon F(x^5)\right).
\end{align}
With
\begin{align}
\exp\left(-\int G_5(x^5) dx^5\right) = \frac{1}{\sqrt{a_1 a_2 a_3 a_4 a_6}}, \quad a_5 = a_4.
\end{align}
Then, \(\tilde{\alpha}_4\) is reduced to
\begin{align}
\tilde{\alpha}_4(x^5, x^6) = \left(e^{-\varepsilon F(x^5)} \tilde{\phi}\right)^* \left(e^{-\varepsilon F(x^5)} \tilde{\phi}\right).
\end{align}
The localization condition requires convergence of the integral of \(\tilde{\alpha}_4\). When \(F\) is a real function, the localization condition can be written as
\begin{align}
\int \exp\left(-2 \varepsilon F(x^5)\right) \tilde{\phi} \tilde{\phi}^* dx^5 dx^6.
\end{align}

Following the field transformation described above, the integrated function in the localization condition can be decomposed into two parts. Here, the field $\tilde{\phi}$ satisfies the Schr\"{o}dinger-like equation, and the effective potential is determined by the warp factor, representing the contribution of the minimal coupling between fermions and gravity. Furthermore, $e^{-2 \varepsilon F(x^5)}$ depends only on the coupling term $F$ rather than the warp (conformal) factor.

The minimal coupling to gravity can be analyzed independently. Thus, this allows us to focus on the geometry of spacetime. For example, one can study the difference in localization behavior under different topologies of spacetime. In contrast, the contributions of other interactions can be analyzed independently. Consequently, the localization condition can work under relatively relaxed conditions, and a universal localization mechanism can be constructed.

Another outcome of conformally transforming the field is that it restores the equations of motion to their flat spacetime form. With the relationship
\begin{align}
{\partial} _5 {\phi}_{ij} = \Upsilon {\partial} _5\tilde{\phi}_{ij} - {H}\tilde{\phi}_{ij},
\end{align}
this allows Eq.~\eqref{h5} to be further simplified to
\begin{subequations}\label{ba4a6}
\begin{align}
m_1\tilde{\phi}_{12} &= -\partial _5\tilde{\phi}_{11} - i b\partial _6\tilde{\phi}_{11},
\\
m_1\tilde{\phi}_{11} &= +\partial _5\tilde{\phi}_{12} - i b\partial _6\tilde{\phi}_{12},
\\
m_2\tilde{\phi}_{22} &= -\partial _5\tilde{\phi}_{21} + i b\partial _6\tilde{\phi}_{21},
\\
m_2\tilde{\phi}_{21} &= +\partial _5\tilde{\phi}_{22} + i b\partial _6\tilde{\phi}_{22},
\end{align}
\end{subequations}
with the expression for \( b\left( x^5 \right) \) given by
\begin{align}
b\left( x^5 \right) = a_4a_{6}^{-1}.
\end{align}
It is worth noting that the warp factor $a_4a_6^{-1}$ appears in pairs. This indicates that the bulk's overall conformal transformation does not affect the form of the equations. It is also important to note that just like the asymptotically AdS bulk and the flat bulk, the localization of fermions cannot be achieved through minimal coupling to gravity. Eq.~\eqref{ba4a6} can subsequently be transformed into a set of independent second-order differential equations
\begin{subequations}
\begin{align}
{m_1}^2\tilde{\phi}_{11}&=-\partial _5\partial _5\tilde{\phi}_{11}-b^2\partial _6\partial _6\tilde{\phi}_{11}-i\partial _5b\partial _6\tilde{\phi}_{11},
\\
{m_1}^2\tilde{\phi}_{12}&=-\partial _5\partial _5\tilde{\phi}_{12}-b^2\partial _6\partial _6\tilde{\phi}_{12}+i\partial _5b\partial _6\tilde{\phi}_{12},
\\
{m_2}^2\tilde{\phi}_{21}&=-\partial _5\partial _5\tilde{\phi}_{21}-b^2\partial _6\partial _6\tilde{\phi}_{21}+i\partial _5b\partial _6\tilde{\phi}_{21}.
\\
{m_2}^2\tilde{\phi}_{22}&=-\partial _5\partial _5\tilde{\phi}_{22}-b^2\partial _6\partial _6\tilde{\phi}_{22}-i\partial _5b\partial _6\tilde{\phi}_{22}.
\end{align}
\end{subequations}

In polar coordinates $x^6 = \theta$, the field function $\tilde{\phi}$ must satisfy a periodic condition. The field function can be decomposed into the following variables
\begin{align}
\tilde{\phi} = u_{r}(r)u_{\theta}(\theta),
\end{align}
with $u_{\theta}(\theta) = e^{i l_6 \theta}$. Upon separation of variables, the equations of motion take the form
\begin{subequations}\label{Schrödinger-like_equation}
\begin{align}
{m_1}^2 u_{11r}&=-\partial _5\partial _5u_{11r} +V_{11}u_{11r},
\\
{m_1}^2 u_{12r}&=-\partial _5\partial _5u_{12r} +V_{12}u_{12r},
\\
{m_2}^2 u_{21r}&=-\partial _5\partial _5u_{21r} +V_{21}u_{21r},
\\
{m_2}^2 u_{22r}&=-\partial _5\partial _5u_{22r} +V_{22}u_{22r},
\end{align}
\end{subequations}
with the corresponding potentials given by
\begin{subequations}
\begin{align}
V_{11}(r)=V_{22}(r)={l_6}^2b^2(r) +l_6\partial _r b(r),
\\
V_{12}(r)=V_{21}(r)={l_6}^2b^2(r) -l_6\partial _r b(r).
\end{align}
\end{subequations}

 {If spacetime is conformally flat, we can find a coordinate such that $a_4=a_6$, thereby implying that $b=1$.} \footnote{ {In the next section, we will see that for two conformally flat topologies in six dimensions, we may not always make the choice of the coordinate systems such that \( a_4 = a_6 \). Therefore, we cannot assert that the spinor field propagates freely along the extra dimensions. The shape of the effective potential will depend on the choice of coordinates, but the judgment of whether the spinor field is localized is independent of the coordinate choice. In more specific cases, we need to further clarify the topology of spacetime.}} In this case, the equations become simplified, and the four two-component fermions satisfy
\begin{align}\label{m2pd5pd6}
m^{2} \tilde{\phi} = -\partial _5\partial _5\tilde{\phi} - \partial _6\partial _6 \tilde{\phi}.
\end{align}
The above equation shares a similar form with the Klein-Gordon equation in flat spacetime, so we have
\begin{align}\label{solution_phi}
\tilde{\phi}_1 = e^{i l_5 x^5}e^{i l_6 x^6},
\end{align}
where $l_5$ and $l_6$ represent the quantized momenta associated with the fermion's movement in the extra dimensions. This yields the energy-momentum relation
\begin{align}
m^2 = l_5^2 + l_6^2.
\end{align}
If $l_5$ and $l_6$ are real, then the localization condition depends only on the integral of the background scalar field:
\begin{align}\label{localization_condition_efphi}
\int e^{-2\varepsilon F(x^5)} d x^5.
\end{align}

In the present section, the condition has been derived for the localization of a bulk fermion in a conformal flat spacetime, analogous to the normalization condition of the Schr\"{o}dinger bound state wave function. Although the asymptotic AdS spacetime helps to localize gravitation, scalar, and vector fields, it proves ineffective for localizing spinor fields, since the asymptotic AdS and flat bulk yield equivalent localization conditions. Consequently, additional interactions or non-minimal couplings between fermionic fields and gravity or background fields are requisite in these extra-dimensional geometries. With interactions present, the aforementioned criterion is augmented by an integral, the nature of which is dictated by the coupling term $\varepsilon  \bar{\Psi} \Gamma^{M} \partial_M {F(\phi, R, R^{\mu\nu}R_{\mu\nu},\cdots)} \Psi$. This criterion is commonly observed in the localization of the braneworld model. The subsequent section will delve into three common extra-dimensional topologies that arise in the $6$-dimensional braneworld scenario.

\section{Examples of $6$-dimensional models}\label{Examples}
 {Remarkably, such a metric form \eqref{line element_x5} can correspond to a variety of topological configurations. In this section, our focus shifts to exploring whether there exists a spacetime structure wherein the spinor field can achieve localization via minimal coupling to gravity. Additionally, we aim to elucidate the tensor coupling mechanism, investigating the terms that can serve as specific interactions for the localization mechanism, given by $F(\phi, R, R^{\mu\nu}R_{\mu\nu}, \ldots)$.}

\subsection{$\mathcal{M}_4\times \mathcal{S}_2$ topology of spacetime}
In the first case, we assume that spacetime is $\mathcal{M}_4\times\mathcal{S}_2$, where $\mathcal{M}_4$ is the $4$-dimensional Minkowski spacetime and $\mathcal{S}_2$ is a two-dimensional compact space. As long as the extra dimensions are compact and continuous, the field function yields a finite integral. Therefore, the localization problem in compact extra dimensions can be naturally solved. However, large-scale compact extra dimensions may also bring some problems. For example, the ADD model \cite{ArkaniHamed:1998rs} with too large extra dimensions will cause non-recoverable Newtonian gravity on the brane. To address this issue, we can consider the possibility of curving extra dimensions, like the RS model. Further, we can even consider constructing brane solutions with compact and warped extra dimensions. Despite the no-go theorem stating that \cite{Gibbons:2000tf, Leblond:2001mr, Leblond:2001xr}, generating a smooth brane on a compact extra-dimensional circular ring using a canonical scalar field in the context of $5$-dimensional general relativity is not possible. However, this limitation does not preclude the possibility in higher dimensions.

\subsection{$\mathcal{M}_4\times {\mathcal{R}_1} \times \mathcal{S}_1$ topology of spacetime}
The second case is that there is both a compact and a non-compact extra dimension.
\begin{figure}[H]
\centering
\includegraphics[width=2.5in]{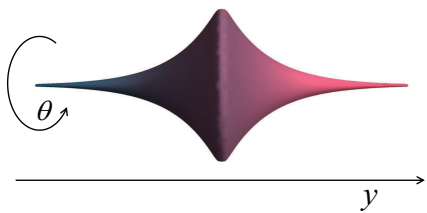}
\caption{The profile of the extra dimensions described by the line element $ds^2_{\text{extra}}=dy^2+b^2(y)R_0^2d\theta^2$, where $y\in(-\infty, \infty)$. In Ref.~\cite{Wan:2020smy}, this transverse space geometry is introduced to achieve the localization of free $U(1)$ gauge field.}
\label{geometric images of extra dimension}
\end{figure}
Assuming the bulk is conformally flat, we can express the spacetime metric as
\begin{eqnarray}
ds^2 = a^2(z)(\eta_{\mu\nu} dx^\mu dx^\nu + dz^2 + d\Theta^2),
\label{metric}
\end{eqnarray}
where, $\eta_{\mu\nu}$ is the four-dimensional Minkowski metric, and $a(z)$ is a warp factor with $Z_2$ symmetry that depends on the extra dimension $z$, with $z \in (-\infty, \infty)$. In this case, the compact radius $R_0$ of the second extra dimension is absorbed into the coordinate $d\Theta = R_0 d \theta$. The localization condition \eqref{localization_condition_efphi} can also be applied for this case. Because the sixth dimension is compact, the field function, which follows the direction $x^6=\Theta$, must satisfy periodic boundary conditions, thereby requiring $l_6$ to be real. On the other hand, to ensure a non-divergent solution at the boundary of the fifth dimension, $l_5$ must be real. This ensures that we have $m^2=l_5^2 + l_6^2 \ge 0$, and hence excludes the existence of tachyon states. The integral with respect to $x^6$ is convergent. Thus, in this scenario, the localization condition simplifies to
\begin{align}\label{localization_condition_efphi2}
\int e^{-2 \varepsilon F(z)}  d z < \infty.
\end{align}

If one only considers the minimal coupling between the fermion and gravity, namely \( F=0 \), it becomes evident that Eq.~\eqref{localization_condition_efphi2} cannot be satisfied. Indeed, Eq.~\eqref{localization_condition_efphi2} necessitates an asymptotic behavior for the coupling term \( e^{-2\varepsilon F(z)} \) that must decline more rapidly than \( 1/|z| \) as \( |z| \rightarrow \infty \). The introduction of a non-vanishing coupling \( F \) is necessary. Exploring new mechanisms with a coupling that satisfy the above condition is an important topic in the context of braneworlds.

Assuming that the spacetime structure is governed by the dynamical scalar field \( \phi \), the warp factor dictates the asymptotic behavior of the background dynamical field through the field equations. Consequently, one must first establish the dynamics of spacetime. It is generally assumed, in general relativity, that a thick brane with \( Z_2 \) symmetry emerges from a background dynamical field \( \phi(z) \) with the action
\begin{eqnarray}\label{action}
S = \frac{{M^4}}{2} \int d^6 x \sqrt{-g} \left(R+\mathcal{L}_m\right),
\label{1}
\end{eqnarray}
where, \( M \) denotes the fundamental scale of the theory, \( g \) is the determinant of the metric \( g_{MN} \), \( R \) is the scalar curvature, and \( \mathcal{L}_m \) is, given the Lagrangian of the background scalar field \( \phi \):
\begin{equation}\label{Lagrangian of the background scalar}
\mathcal{L}_m =-\frac{1}{2}g^{MN}{\partial_M}\phi \partial_N\phi-V_\Lambda(\phi ),
\end{equation}
associated with the scalar potential
\begin{eqnarray}
V_\Lambda(\phi)&=&V(\phi)+\Lambda.
\end{eqnarray}
In this context, \( \Lambda \) symbolizes the cosmological constant and \( V(\phi(z\rightarrow \pm\infty)) \rightarrow 0 \).

In the context of braneworld models, the introduction of a non-zero cosmological constant can lead to an asymptotically AdS spacetime, which has desirable properties for the localization of a scalar field and gravity. This asymptotic behavior is commonly assumed in many braneworld models. In accordance with the Einstein field equations, the following relationship exists between the background scalar field and the curvature factor:
\begin{eqnarray}
\frac{2(D-2) a^{\prime}(z)^2}{a(z)^2}-\frac{(D-2) a^{\prime \prime}(z)}{a(z)}=\left(\phi'(z)\right)^2,
\end{eqnarray}
where \( D \) is the dimension of the spacetime. For the asymptotically AdS bulk spacetime under consideration, we should have:
\begin{eqnarray}\label{warp_factor_asymptotically_behavior}
	a(z\rightarrow \pm\infty)&\rightarrow& \frac{1}{|z|}, \\
    \phi'(z\rightarrow \pm\infty) &\rightarrow& 0.
\end{eqnarray}
Moreover, the asymptotic behavior of the background scalar field is characterized by one of the following three cases:
\begin{eqnarray}
   \text{case I:~~~} |\phi(z)| &\rightarrow& \infty~~~\text{and~~~} \frac{|\phi(z)|}{ \text{log} |z|} \rightarrow 0, \label{phiAsymptoticallyBehavior1} \\
   \text{case II:~~~~\,} \phi(z) &\rightarrow& v_{\pm} , \label{phiAsymptoticallyBehavior2}\\
   \text{case III:~~~~\,} \phi(z) &\rightarrow& 0 ,  \label{phiAsymptoticallyBehavior3}
\end{eqnarray}
at the boundary \( |z| \rightarrow  \pm\infty \),
where \( v_+ \) and \( v_- \) are constants, respectively, which usually correspond to local stable points in the system, taking into account both gravity and the scalar potential.

If the scalar potential \( V(\phi)|_{\phi\rightarrow \pm \infty} \rightarrow V_{\text{min}} = 0 \), where \( V_{\text{min}} \) is the minimum value of \( V(\phi) \), this corresponds to case I when \( {|\phi(z)|} \rightarrow \text{log}(\text{log}|z|) \).
If we consider \( F=\phi \), the localization condition \eqref{localization_condition_efphi2} requires that
\begin{eqnarray}
   \log |z|/\phi(z) \rightarrow 0  ~~~\text{at}~~~ z \rightarrow \pm \infty,  \label{condition1}
\end{eqnarray}
which conflicts with the asymptotic behavior of the background scalar \eqref{phiAsymptoticallyBehavior1}.
Nevertheless, considering \( F=\phi^n \) for \( n>1 \) may provide a viable localization mechanism, since the localization condition corresponding to \eqref{localization_condition_efphi2} for this case is not in conflict with \eqref{phiAsymptoticallyBehavior1}.

For the other two cases of \( \phi(z \rightarrow \pm \infty) \rightarrow v_{\pm} \) and \( \phi(z \rightarrow \pm \infty) \rightarrow 0 \), we can refer to the two explicit solutions given in Ref. \cite{Wan:2020smy} as examples. Now we show an example for \( \phi \rightarrow v_{\pm} \). The scalar potential, the scalar field, the warp factors, and the \( 5 \)-dimensional cosmological constant are given by \cite{Wan:2020smy}
\begin{eqnarray}
V (\phi)&=&
\frac{k^{2} v^{2}}{2}+\frac{5}{18} k^{2} v^{4}
-\left(k^{2}+\frac{5 k^{2} v^{2}}{8}\right) \phi^{2}+\left(\frac{5 k^{2}}{12}+\frac{k^{2}}{2 v^{2}}\right) \phi^{4}-\frac{5 k^{2} }{72 v^{2}}\phi^{6}, \\
\phi (y)&=&v \operatorname{tanh}(k y), \label{BackgroundScalarField1} \\
a(y) &=& b(y)=\mathrm{e}^{-\frac{1}{24} v^{2} \tanh^{2} (k y)} \text{sech}^{\frac{v^{2}}{6}}(k y),\\
\Lambda&=&-\frac{5}{18} k^{2} v^{4}.
\end{eqnarray}
Here \( v \) is a dimensionless parameter, \( k \) is a fundamental energy scale with dimension \([k]=L^{-1}\), and \( 1/k \) stands for the thickness of the brane. When converted to the \( z \) coordinate with \( dz = a^{-1}(y) dy \), we have
\begin{subequations}
\begin{eqnarray}
\phi(z) &\rightarrow& v\frac{-1+k^2 z^2}{1+k^2 z^2} \rightarrow v   \quad \text{when} ~ z \rightarrow \infty, \\
\phi(z) &\rightarrow& v\frac{1-k^2 z^2}{1+k^2 z^2} \rightarrow -v   \quad \text{when} ~ z \rightarrow -\infty.
\end{eqnarray}
\end{subequations}
%The profile of the background scalar field \( \phi \) is plotted in Fig.~\ref{figBackgroundScalarField1}.
In this scenario, the function \( F(\phi) \) needs to diverge when \( \phi\rightarrow \pm v \) in order to satisfy the localization condition.
%\begin{figure}[H]
%\center{
%\includegraphics[width=2.5in]{solution1.eps}
%}
%\caption{ The profile of the background scalar field $\phi$ given in Eq.~\eqref{BackgroundScalarField1}, when $z\rightarrow \pm \infty$, $\phi \rightarrow \pm v$.
%}
%\label{figBackgroundScalarField1}
%\end{figure}

For the case where $\phi \rightarrow 0$, a brane solution can be found if the scalar potential $V(\phi)$ and the cosmological constant $\Lambda$ are given by
\begin{eqnarray}
V(\phi) &=&
\left(\frac{k^{2}}{2}+\frac{5 k^{2} v^{2}}{24}\right) \phi^{2}
-\left(\frac{5 k^{2}}{24}+\frac{k^{2}}{2 v^{2}}\right) \phi^{4}
+\frac{5 k^{2}}{72 v^{2}}\phi^{6}, \label{Scalarpotential2}\\
\Lambda &=& -\frac{5k^2 v^4}{72}.
\end{eqnarray}
The expressions for the scalar field and the warp factors are taken from Ref.~\cite{Wan:2020smy} and are presented as follows:
\begin{eqnarray}
    \phi(y) &=& v~\text{sech}(k y), \label{BackgroundScalarField2}\\
	a(y) &=& b(y)=\mathrm{e}^{\frac{1}{24} v^2\tanh ^2(k y)}{{\text{sech}^{\frac{v^2}{12}} (k y)}}. \label{warpfactors2}
\end{eqnarray}
%The profile of the background scalar field $\phi$ is illustrated in Fig.~\ref{figBackgroundScalarField2}.
When converted to the $z$ coordinate, the field $\phi(z)$ evolves as follows,
\begin{eqnarray}
\phi(z) \rightarrow v\frac{2 k z}{1+k^2 z^2} \rightarrow 2v\frac{1}{k z} \rightarrow 0,  \quad \text{when} ~ z \rightarrow \pm \infty.
\end{eqnarray}
This serves as a specific example. In fact, as long as $\phi(z) \rightarrow z^{-\alpha}$ with $\alpha>0$, the spacetime is asymptotically AdS. For this case, we can also consider the localization mechanism with the coupling function $F(\phi)=\phi^n$. This leads to the integration factor $\exp(-2 \varepsilon \phi^n)$, which tends toward $\exp(-2 \varepsilon |z|^{-n \alpha})$ when $|z|\rightarrow \infty$. For $n\geq0$, where $e^{-2 \varepsilon F(\phi)} \sim 1$ at the boundary, the localization condition cannot be met. Conversely, for $n < 0$, the integration factor $e^{-2 \varepsilon F(\phi)} \rightarrow \exp(-2 \varepsilon |z|^{|n \alpha|})$ at $|z|\rightarrow \infty$, and the localization condition can be satisfied.

%\begin{figure}[H]
%\center{
%\includegraphics[width=2.5in]{solution2.eps}
%}
%\caption{The profile of the background scalar field $\phi$ given in Eq.~\eqref{BackgroundScalarField2}.
%}
%\label{figBackgroundScalarField2}
%\end{figure}

As described earlier, we have transformed the effects of gravity into flat spacetime through field transformation. This lack of coupling prevents the fermion from being localized on the brane. For an asymptotically AdS spacetime, the curved spacetime does not help to localize the matter field. This implies that the introduction of the coupling term is necessary, and the coupling function $F(\phi)$ cannot go to a constant at the boundary. One should carefully choose the form of the coupling $F(\phi)$, such that it still maintains a strong coupling between the fermion and the background scalar field at infinity. We have shown that $F = \phi^n$ is a possible choice of localization scheme. For $\phi(z\rightarrow \pm \infty) \rightarrow \infty$, the localization condition requires $n>1$. Conversely, for $\phi(z\rightarrow \pm \infty) \rightarrow 0$, it requires $n<0$.

If the warp factor is solely a function of one of the extra dimensions, then minimal gravitational coupling fails to distinguish between the high-dimensional left and right chiral massless fermions. This is consistent with the results of the $5$-dimensional Randall-Sundrum-like model.

When $a_4 = a_6$, all the extra-dimensional components of the four $2$-component spinors obtained from the 8-component spinor decomposition satisfy the same equation of motion. In other words, the localization mechanism fails to distinguish between the left- and right-handed chiralities of the fermion. However, for the topology $\mathcal{M}_4\times {\mathcal{R}_2}$, which will be considered subsequently, minimal coupling between fermions and gravity is sufficient to distinguish chirality.

\subsection{$\mathcal{M}_4\times {\mathcal{R}_2}$ topology of spacetime}
\begin{figure}[H]
\center{
\includegraphics[width=2.5in]{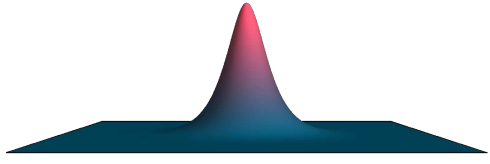}
}
\caption{The profile of the extra dimensions described by the line element (\ref{brane2}).
}
\label{geometric images of extra dimension_2}
\end{figure}

For the topology $\mathcal{M}_4\times {\mathcal{R}_2}$, there are two non-compact extra dimensions. The brane on which we live has a codimension of 2 with respect to the bulk, and the extra dimensions have a non-trivial structure shown in Fig.~\ref{geometric images of extra dimension_2}. We are interested in the bulk spacetime, which is symmetric with respect to the brane. The metric of the spacetime is assumed to be
\begin{align}
ds^2=a_4^2(r)\eta_{\mu\nu} dx^{\mu} dx^{\nu} + a_4^2(r) dr^2 + a_6^2(r) d\theta^2. \label{brane2}
\end{align}
Eq.~\eqref{brane2} describes this metric, which satisfies the requirements of Assumption 4. Under this metric, the extra-dimensional part of the fermion satisfies the Schr\"{o}dinger-like equation given by Eq.~\eqref{Schrödinger-like_equation}.
We further assume that the spacetime is conformally flat. Consequently, the metric can be expressed as
\begin{equation}
ds^2=a_4^2(r)(\eta_{\mu\nu} dx^{\mu}dx^{\nu} + dr^2 + r^2 d\theta^2). \label{ConformalFlatMetric}
\end{equation}
Then we find \( b = a_4 a_6^{-1} = \frac{1}{r} \), and
\begin{subequations}
\begin{align}
V_{11}(r)&=V_{22}(r)=\frac{l_6(l_6-1)}{r^2}, \\
V_{12}(r)&=V_{21}(r)=\frac{l_6(l_6+1)}{r^2}.
\end{align}
\label{PotentialTerms}
\end{subequations}
Eq.~\eqref{PotentialTerms} gives the potential terms under this conformally flat metric. The extra-dimensional parts of the left-handed and right-handed chiral components of the four-dimensional fermions \( \Psi _{1}^{(4)} \) (the analysis for \( \Psi _{2}^{(4)} \) is similar) satisfy the Schr\"{o}dinger-like equations with different effective potentials. Given by
\begin{subequations}
\begin{align}
{m_1}^2 u_{11r} &= -\partial _5\partial _5u_{11r} + \frac{l_6(l_6-1)}{r^2}u_{11r}, \\
{m_1}^2 u_{12r} &= -\partial _5\partial _5u_{12r} + \frac{l_6(l_6+1)}{r^2}u_{12r}.
\end{align}
\label{SchrödingerEquations}
\end{subequations}
This means that in this topology, gravity differentiates between the left-handed and right-handed chiralities of the \(4\)-dimensional fermions. We need to emphasize that here the left- and right-handed chiralities refer to those of the \(4\)-component spinor, not the eight-component spinor. This may lead to a chiral theory on the brane. For an asymptotically AdS spacetime with the topology \(\mathcal{M}_4 \times \mathcal{R}_2\), the localization mechanisms of the minimal gravity coupling and the geometric coupling mentioned earlier cannot ensure the localization of fermions on the brane. The reason is as follows:
\begin{itemize}
  \item In the case of the minimal gravity coupling (\(F = 0\)), we have \(V_{11}(r) \rightarrow 0\), \(V_{12}(r) \rightarrow 0\) at \(r\rightarrow\infty\). This means that \(u_{11r}\) and \(u_{12r}\) are not ``bound states,'' and the localization condition is not satisfied\footnote{We have noticed that there are infinitely degenerate states in the zero mass KK mode. Among these, the case of \(l_6 = 0, 1\) is particularly special, since it leads to the effective potential \(V_{11} = V_{22} = 0\). (For \(l_6 = -1, 0\), \(V_{12} = V_{21} = 0\) as well.) This is the same situation as in Eq.~\eqref{m2pd5pd6}. When \(l_6\) takes other values, a potential barrier rather than a potential well will form at the origin, and the effective potential is a monotonic function of \(r\)}.
  \item For the geometric coupling with \(F(R) = R^n\), the integral factor \(e^{-2 \varepsilon F(R(z))}\) converges to a constant when \(z\rightarrow \pm \infty\) since the scalar curvature \(R\) is a constant at the boundary of the extra dimensions in asymptotically AdS spacetime. Thus, the localization condition cannot be satisfied.
\end{itemize}
Even if the bulk topology changes, it still requires that the kinetic term of the scalar field has no contribution at the boundary of the spacetime in the asymptotically AdS spacetime. Therefore, the coupling with \(F(\phi)=\phi^n\) remains a viable option. Notice that \(V_{11} \rightarrow 0\) and \(V_{22} \rightarrow 0\) when \(r\rightarrow\infty\), \(u_{11}(r)\) and \(u_{12}(r)\) will converge to free wave functions at the boundary. When \(\phi \rightarrow 0\) as \(r \rightarrow \infty\), the localization condition is satisfied for \(n < 0\). Furthermore, we can finely adjust \(F(\phi)\) to satisfy \(z \times e^{-2 \varepsilon F(\phi)}\sim 1\) so that only one chirality of the fermion is localized on the brane.

If the topology of extra dimensions is \(\mathcal{R}_2\) and the bulk is conformally flat, a minimal coupling to gravity without other interactions cannot localize fermions. If the bulk is conformally AdS and a derivative coupling mechanism such as partial derivative coupling with the background scalar field or curvature is introduced, then the two topologies considered in this section also yield similar results for fermion localization. The main difference between the two is that the topology of \(\mathcal{M}_4 \times \mathcal{R}_1 \times \mathcal{S}_1\) does not distinguish between left- and right-handed fermions, whereas the topology of \(\mathcal{M}_4 \times \mathcal{R}_2\) does. Since there is a difference in the effective potentials for the left- and right-handed fermions, the \(4\)-dimensional chiral theory can be restored by fine-tuning the coupling function \(F(\phi)\).

\section{Conclusion and discussion}\label{Conclusions_Outlook}
Due to the constraints imposed by the no-go theorem \cite{Gibbons:2000tf, Leblond:2001mr, Leblond:2001xr}, the $5$-dimensional braneworld model ignores a significant number of potential extra-dimensional topological structures and thick brane solutions. Therefore, these should be considered in higher-dimensional spacetime.  {In circumventing the no-go theorem, two potential avenues emerge: firstly, delving into models with increased dimensions, and secondly, contemplating modifications to gravity.
By introducing a dilaton field, one can construct a high-dimensional spacetime with compact dimensions, and this approach simultaneously circumvents the two conditions for the no-go theorem to hold \cite{DeFelice:2008af}.}

Additionally, for the localization mechanism of matter fields, the most direct approach is to extend the dimensions of spacetime, a technique that is particularly effective for handling other types of matter fields, especially $U(1)$ gauge fields. Previous studies have shown that a $6$-dimensional spacetime can facilitate the localization of $U(1)$ gauge fields through minimal coupling with gravity, a feat unattainable in $5$-dimensional RS-like models \cite{Bajc:1999mh}.

However, the extension of dimensions poses challenges for the localization of fermions, as every two-dimensional increase in the momentum space results in a doubling of the dimensionality of the spinor representation space. It becomes necessary to map the degrees of freedom in the higher-dimensional theory to those in the effective $4$-dimensional theory, with the aim of recovering the $4$-dimensional chiral theory.

Our work builds upon the research detailed in Ref.~\cite{Budinich:2001nh}, which explores the relationship between high-dimensional and $4$-dimensional fermions in the context of braneworlds, using the relationship between Clifford algebra and its subalgebra. Our ongoing research focuses on the localization of fermions in higher dimensions and aims to develop a comprehensive theory to explain their behavior in these scenarios. We explore the representation of spinor fields in even-dimensional spacetimes, derive the equations of motion for spinors in curved spacetime, and furnish common calculations pertinent to braneworld models. Our findings indicate that in a conformally flat, extra-dimensional spacetime, fermions cannot be localized solely through minimal gravitational coupling. Such a result necessitates the consideration of the interaction between background dynamical fields and fermions.

To preserve Lorentz symmetry in higher-dimensional spacetime, and to facilitate the decoupling of the components of the higher-dimensional spinor to obtain a $4$-dimensional effective free field theory, we introduce the $\bar{\Psi} \Gamma^{M}\Gamma^{N}\Gamma^{P}\cdots T_{MNP\cdots} \Psi$ coupling mechanism. For instance, we opt for the partial derivative of a scalar function for constructing a first-order tensor that couples with the spinor field as follows: $\varepsilon \bar{\Psi} \Gamma^{M} \partial_M F(\phi, R, R^{\mu\nu}R_{\mu\nu},\cdots) \Psi$. The inclusion of these coupling terms yields an interaction term $e^{-2 \varepsilon F(z)}$, which appears in the integral of the localization conditions.

For the localization of fermions, it is necessary that the coupling function be sufficiently large as \( z \rightarrow \infty \) to provide a sufficiently strong interaction that localizes the fermion. In an asymptotically AdS bulk, the field function tends toward a constant value as \( z \rightarrow \infty \). Both the background dynamical fields and the background geometry should remain finite at infinity. In such a scenario, we suggest a coupling mechanism of \( F(\phi)=\phi^n \). Localization conditions dictate that the different behavior of the field \( \phi \) imposes different requirements on the value of \( n \). When the bulk topology is \( \mathcal{M}_4\times {\mathcal{R}_1} \times \mathcal{S}_1 \), for \( \phi(z\rightarrow \pm \infty) \rightarrow \infty \), the localization condition requires \( n>1 \), while for \( \phi(z\rightarrow \pm \infty) \rightarrow 0 \), it requires \( n<0 \).

In braneworld theory, it is desirable to consider a model in which left-handed chiral particles can be localized, while right-handed chiral particles cannot be localized. This provides a theoretical basis for the restoration of the standard model fermion chiral theory, and it is also one of the motivations for braneworld models. However, not all models are capable of realizing this concept. Interestingly, the topology of the extra dimensions affects the localization of left- and right-handed chiral fermions. In conformally flat spacetimes, different extra-dimensional topological structures can result in differences in the chirality of fermions. If we only consider the minimal coupling between fermions and gravity and ignore other localization mechanisms, no difference arises between the left and right chiralities of fermions in a bulk with the topology of \( \mathcal{M}_4\times {\mathcal{R}_1} \times \mathcal{S}_1 \). However, the left and right chiralities of fermions are distinguished in \( \mathcal{M}_4\times {\mathcal{R}_2} \).

The complex geometry of higher-dimensional spacetimes, an increased number of degrees of freedom of the matter field, and the possibility of coupling in localization mechanisms result in a wide range of potential braneworld models. However, if we aim for the models to be straightforward, we face various difficulties and contradictions. In order to obtain a $4$-dimensional chiral theory, we suggest further study of new localization mechanisms for the topology \( \mathcal{M}_4\times {\mathcal{R}_2} \) in the context of conformally flat extra dimensions. In contrast, the topology of \( \mathcal{M}_4\times {\mathcal{R}_1} \times \mathcal{S}_1 \) allows the localization of free vector fields that have minimal coupling to gravity, while the topology of \( \mathcal{M}_4\times {\mathcal{R}_2} \) does not \cite{Wan:2020smy}. Therefore, the \( U(1) \) gauge field and the spinor field have distinct topological requirements for extra dimensions.

One approach could be to retain the benefits of both topological structures in a higher-dimensional spacetime, for instance, in \( \mathcal{M}_4\times {\mathcal{R}_2} \times \mathcal{S}_1 \). Through minimal coupling with gravity, the question of whether the left and right chiralities of fermions can be distinguished or if the \( U(1) \) gauge field can be localized simultaneously warrants further research. Additional investigation is required to develop a higher-dimensional braneworld theory that is both concise and self-consistent, without being confined to 5 or 6 dimensions.

\appendix
%\section{Clifford algebra $\mathbb{C} \ell(2n+1,1)$ and representations of the Gamma matrix}\label{Clifford algebra}
\section{Clifford algebra $\mathbb{C} \ell(2n+1,1)$}
Given a $(2n+2)$-dimensional complex space $\mathbb{C}^{N}$,
the $(2n+2)$-dimensional Clifford algebra $\mathbb{C} \ell(2n+1,1)$ generated by Dirac matrices $\Gamma_{A}$ $(A=0,1,2,\cdots,2n+1)$ is defined by
\begin{equation}
\left\{\Gamma_{A}, \Gamma_{B}\right\}=2 \eta_{A B} I_{2n+2},
\end{equation}
where $I_{2n+2}$ represents a $(2n+2)\times (2n+2)$ unit matrix and $\eta_{AB}=\text{diag} (-,+,+,\cdots,+)$ is the Minkowski metric.
The above definition can be generalized to the scenario of a curved spacetime
\begin{equation}
\left\{ E_M^{~~A} \Gamma_{A}, E_N^{~~B} \Gamma_{B}\right\}=2 g_{MN} I_{2n+2}
\end{equation}
with
\begin{equation}
g_{M N}=E_{M}^{~~A} E_{N}^{~~B} \eta_{A B}.
\end{equation}
In this paper,  $E_{M}^{~~A}$ are the vielbein fields, $A,B,\cdots$ and $M,N,\cdots$ denoting the ``Lorentz index'' and  ``spacetime index'', respectively.

\section{Representations of the Gamma matrix}
Denoting the generators of $\mathbb{C} \ell(2 n+1,1)$ and of $\mathbb{C} \ell(2 n-1,1)$ as $\Gamma_{A}$ and $\gamma_{a}$ respectively, we can express the $2 n+2$ generators $\Gamma_{A}$ and the ``volume element'' $\Gamma_{2 n+3}$ as follows:
\begin{align}
\Gamma_{0}&=\sigma_{1} \otimes \gamma_{0},
\nonumber \\
\Gamma_{A}
&=\sigma_{1} \otimes \gamma_{a}, \quad (A=a=1,2, \cdots, 2 n-1)       \nonumber \\
\Gamma_{2n}&=\sigma_{1} \otimes \gamma_{2n+1},                        \nonumber \\
 \Gamma_{2n+1}&=\sigma_{2} \otimes I_{2^{n}},                         \nonumber \\
\Gamma_{2 n+3}&\equiv (-i)^{n} \prod_{A=0}^{2n+1}\Gamma_A
=\sigma_{3} \otimes I_{2^{n}} =\left(
                                  \begin{array}{cc}
                                    I_{2^{n}} & 0 \\
                                    0 & -I_{2^{n}} \\
                                  \end{array}
                                \right).
\end{align}
Here, $\sigma_{i}$ are the Pauli matrices:
\begin{equation}
\sigma_{1}=\left(\begin{array}{ll}
0 & 1 \\
1 & 0
\end{array}\right), \quad \sigma_{2}=\left(\begin{array}{cc}
0 & -i \\
i & 0
\end{array}\right), \quad \sigma_{3}=\left(\begin{array}{cc}
1 & 0 \\
0 & -1
\end{array}\right).
\end{equation}
And $\gamma_{2n+1}$ is defined as
\begin{eqnarray}
\gamma_{2n+1}\equiv(-i)^{(n-1)}\prod_{a=0}^{2n-1}\gamma_a=\sigma_{3} \otimes I_{2^{n-1}}.
\end{eqnarray}
The volume element $\Gamma_{2 n+3}$ anticommutes with all the other $2n+2$ Gamma matrices $\Gamma_A$, and all of these $2n+3$ Gamma matrices generate the $(2n+3)$-dimensional Clifford algebra $\mathbb{C} \ell(2 n+2,1)$. The volume element can also generate a projection operator to obtain the subalgebra $\mathbb{C} \ell_0(2 n+1,1)$ of $\mathbb{C} \ell(2 n+1,1)$, where the subscript ``0'' indicates the subalgebra. In this framework,  the chiral projection operator can be written as:
\begin{align}\label{projection operator 4D}
P_L&=\frac{I_{2^{n+1}} +\Gamma_{2n+3}}{2}=\left(\begin{array}{cc}
I_{2^n} & 0 \\
0 & 0_{2^n}
\end{array}\right),  \\
P_R&=\frac{I_{2^{n+1}}-\Gamma_{2n+3}}{2}=\left(\begin{array}{cc}
0_{2^{n} } & 0 \\
0 & I_{2^{n} }
\end{array}\right),
\end{align}
These projection operators $P_L$ and $P_R$ act to decompose a $2^{n+1}$-component Dirac spinor $\Psi^{(2\iota)}$ into two $2^n$-component Weyl spinors $\Psi_1^{(\iota)}$ and $\Psi_2^{(\iota)}$ as follows:
\begin{equation}\label{Psi_left_right}
\begin{array}{l}
\Psi^{(2\iota)}_L=P_{L} \Psi^{(2\iota)}=\left(\begin{array}{c}
\Psi_{1}^{(\iota)} \\
0
\end{array}\right),~\text{and} ~
\Psi^{(2\iota)}_R=P_{R} \Psi^{(2\iota)}=\left(\begin{array}{c}
0 \\
\Psi_{2}^{(\iota)}
\end{array}\right),
\end{array}
\end{equation}
where $2\iota=2^{n+1}$ represents the number of spinor components.
$\Psi_1^{(\iota)}$ and $\Psi_2^{(\iota)}$ are vectors of the representation space of the even subalgebra $\mathbb{C} \ell_0(2 n+1,1)$ of $\mathbb{C} \ell(2 n+1,1)$ and their non-zero components span $\iota$-dimensional Weyl spinor spaces.

 {For $D  = 2 ~ mod ~4$ each Weyl representation is its own conjugate. Majorana condition ($\zeta=B^* \zeta^*=B^* B \zeta$) is possible if $D  = 0, 1, 2, 3, \text{or}~4~(mod~ 8)$. Therefore, one might obtain Majorana-Weyl spinors (For example, in the 10-dimensional case). For a more detailed discussion, one can refer to the Table B.1 in Ref. \cite{Polchinski:1998rr}.}

%For  $d = 0 ~ mod ~ 4$ each Weyl representation is conjugate to the other. Thus in $ d = 4$ we can designate the representations as a $2$-component Weyl spinor and its conjugate.

\subsection{Weyl representation for $\mathbb{C} \ell(3,1)$ and $\mathbb{C} \ell(5,1)$}
As an example, one concrete form of the Gamma matrices for $\mathbb{C} \ell(3,1)$  can be written as
\begin{align}\label{gamma metric}
	\gamma_{0}=i \sigma_2 \otimes I_{2}=
	\left(\begin{array}{ll}
		0 & I_{2} \\
		-I_{2} & 0
	\end{array}\right), \quad
	\gamma_{i}
	=
	\sigma_1 \otimes \sigma_i
	=
	\left(\begin{array}{cc}
		0 & \sigma_{i} \\
		\sigma_{i} & 0
	\end{array}\right), ~ i=1,2,3
\end{align}
and
\begin{align}\label{gamma5}
	\gamma_5
	={-i} \gamma_0\gamma_1\gamma_2\gamma_3
	=\sigma_3 \otimes I_2
	=
	\left(
	\begin{array}{cc}
		I_2 & 0 \\
		0 & -I_2
	\end{array}
	\right).
\end{align}
In this convention, the matrix $\gamma^0$ is anti-hermitian, while the other gamma matrices are hermitian ($\gamma_5$ is also hermitian). The Dirac operator acts as $\gamma^{\mu}\partial_{\mu}\psi=m\psi$, and it is easy to see that $\gamma^{5}\psi_{L,R}=\pm \psi_{L,R}$ from Eq.~\eqref{gamma5} and Eq.~\eqref{Psi_left_right}. It is the direct sum of the two irreducible representations of the Lorentz group: $\left(\frac{1}{2}, 0\right) \oplus\left(0, \frac{1}{2}\right)$. The Lorentz generators $S^{\mu\nu}=\frac{i}{4}\left[\gamma^{\mu}, \gamma^{\nu}\right]$ are block diagonal. Under an infinitesimal Lorentz transformation
\begin{equation}
	\psi \rightarrow \psi+\frac{1}{2}\left(\begin{array}{cc}
		\left(i \theta_{i}-\beta_{i}\right) \sigma_{i} & 0\\
		0& \left(i \theta_{i}+\beta_{i}\right) \sigma_{i}
	\end{array}\right) \psi,
\end{equation}
where $\theta_{i}$ and $\beta_{i}$ represent the rotation angle for rotation and the boost respectively.
In this basis, a Dirac spinor is a doublet of a left and a right-handed Weyl spinor:
\begin{eqnarray}
	\psi=\left(\begin{array}{c}
		\psi_{L} \\
		\psi_{R}
	\end{array}\right).
\end{eqnarray}
Here, left-handed and right-handed refer to the $\left(\frac{1}{2}, 0\right)$ or $\left(0, \frac{1}{2}\right)$ representations of the Lorentz group. And the Lorentz scalar is structured as
\begin{equation}
	\bar{\psi} \psi= \bar{\psi}_R \psi_L + \bar{\psi}_L \psi_R.
\end{equation}
The above representation is called the chiral representation.

Further, we can generalize this to a $6$-dimensional representation using $\Gamma_{A}^{\{1\}}$ in Eq.~\eqref{to_higher_dimensional_gamma} as
\begin{align}\label{6d_gamma}
	\Gamma^{{\mu}}=\left(\begin{array}{cc}
		0 & \gamma^{{\mu}} \\
		\gamma^{{\mu}} & 0
	\end{array}\right),
	\quad
	\Gamma^{{5}}=\left(\begin{array}{cc}
		0 & \gamma^{5} \\
		\gamma^{5} & 0
	\end{array}\right),
	\quad
	\Gamma^{{6}}=\left(\begin{array}{cc}
		0 & -i \\
		i & 0
	\end{array}\right),
	\quad
	\Gamma^{{7}}=\left(\begin{array}{cc}
		I_4 & 0 \\
		0 & -I_4
	\end{array}\right).
\end{align}
We will use this representation to investigate the localization of the $6$-dimensional spinor field. With this representation, it becomes convenient to decompose the left and right chiral spinors, showing that the decomposed $4$-dimensional part of the left or right chiral component corresponds to the matter field in the $4$-dimensional effective theory.

\subsection{The representations and the transformations between them}
In fact, including the Gamma matrices constructed above (denoted as $\Gamma^{\{1\}}_A$),
we can also obtain four representations of the Gamma matrices \cite{Budinich:2001nh}:
\begin{align}\label{to_higher_dimensional_gamma}
\begin{array}{llll}
\Gamma_{a}^{\{0\}}=1 \otimes \gamma_{a}, &\Gamma_{2 n+1}^{\{0\}}=\sigma_{1} \otimes \gamma_{2 n+1}, & \Gamma_{2 n+2}^{\{0\}}=\sigma_{2} \otimes \gamma_{2 n+1}, & \Gamma_{2 n+3}^{\{0\}}=\sigma_{3} \otimes \gamma_{2 n+1}; \\
\Gamma_{a}^{\{1\}}=\sigma_{1} \otimes \gamma_{a}, &\Gamma_{2 n+1}^{\{1\}}=\sigma_{1} \otimes \gamma_{2 n+1}, & \Gamma_{2 n+2}^{\{1\}}=\sigma_{2} \otimes 1, & \Gamma_{2 n+3}^{\{1\}}=\sigma_{3} \otimes 1; \\
\Gamma_{a}^{\{2\}}=\sigma_{2} \otimes \gamma_{a}, &\Gamma_{2 n+1}^{\{2\}}=\sigma_{2} \otimes \gamma_{2 n+1}, & \Gamma_{2 n+2}^{\{2\}}=\sigma_{1} \otimes 1, & \Gamma_{2 n+3}^{\{2\}}=\sigma_{3} \otimes 1; \\
\Gamma_{a}^{\{3\}}=\sigma_{3} \otimes \gamma_{a}, &\Gamma_{2 n+1}^{\{3\}}=\sigma_{3} \otimes \gamma_{2 n+1}, & \Gamma_{2 n+2}^{\{3\}}=\sigma_{2} \otimes 1, & \Gamma_{2 n+3}^{\{3\}}=\sigma_{1} \otimes 1,
\end{array}
\end{align}
where the superscript of the Gamma matrices $\{i\}$ serves to mark the $i$th representation of the Gamma matrices.

The transformation between the $0$th and $i$th representations is given by
\begin{equation}\label{gamma}
U_{j} \Gamma_{A}^{\{0\}} U_{j}^{-1}=\Gamma_{A}^{\{j\}}, \quad (A=1,2, \cdots, 2 n+2, \quad j=1,2,3) ,
\end{equation}
where
\begin{equation}
U_{j}:=1_{2} \otimes P_L+\sigma_{j} \otimes P_R=U_{j}^{-1},
\end{equation}
and the transformation between the corresponding spinors is
\begin{equation}\label{Dirac_representation_transformation}
U_{j} \Psi^{\{0\}}=\Psi^{\{j\}}.
\end{equation}
This transformation leaves the form of the Dirac equation invariant. Under the Dirac representation \(\Gamma^{\{0\}}\), the Dirac equation takes the form
\begin{equation}
\left(\Gamma_{M}^{\{0\}} D^{M}-m\right) \Psi^{\{0\}} =0,
\end{equation}
where
\begin{equation}
D_{M}=\partial_{M}+\Omega_{M}
\end{equation}
and
\begin{equation}
\Omega_{M}=\frac{1}{4} \Omega_{M}^{A B} \Gamma_{A}^{\{0\}} \Gamma_{B}^{\{0\}}
\end{equation}
with the spin connection  $\Omega_{M}^{A B}$ defined as
\begin{align}
\begin{aligned}\label{spinconnection}
\Omega_{M}^{A B} &=\frac{1}{2} E^{N A}\left(\partial_{M} E_{N}^{B}-\partial_{N} E_{M}^{B}\right)-\frac{1}{2} E^{N B}\left(\partial_{M} E_{N}^{A}-\partial_{N} E_{M}^{A}\right) \\
&-\frac{1}{2} E^{P A} E^{Q B}\left(\partial_{P} E_{Q C}-\partial_{Q} E_{P C}\right) E_{M}^{C}.
\end{aligned}
\end{align}
The Dirac equation in terms of \(\Gamma^{\{j\}}\) transforms as:
\begin{equation}
 \left(\Gamma_{M}^{\{j\}}  D^{M}-m\right)  \Psi^{\{j\}}=0
\end{equation}
with \(\Omega_{M} = \frac{1}{4} \Omega_{M}^{AB} \Gamma_{A}^{\{j\}} \Gamma_{B}^{\{j\}}\),
which shows that a change in representation does not affect the dynamics of the spinor.

In fact, a different representation means a different set of bases of the spinor space. Let
\begin{equation}\label{Chiral_decomposition_of_spinor}
\Psi=\left(\begin{array}{l}
\Psi_{1}^{(\iota)} \\
\Psi_{2}^{(\iota)}
\end{array}\right),
\end{equation}
be a \(2^{n+1}\)-component spinor associated with \(\mathbb{C} \ell(2 n+2)\). It can be easily seen that the \(2^{n}\)-component spinors \(\Psi_{1}^{(\iota)}\) and \(\Psi_{2}^{(\iota)}\) may be
\begin{itemize}
  \item \(\mathbb{C} \ell(2 n)\) - Dirac spinors (\(\Psi^{\{0\}}\)) for \(\Gamma_{A}^{\{1\}}\),
  \item \(\mathbb{C} \ell_{0}(2 n+2)\) - Weyl spinors (\(\Psi^{\{1\}}\) or \(\Psi^{\{2\}}\)) for \(\Gamma_{A}^{\{1\}}\) or \(\Gamma_{A}^{\{2\}}\),
  \item \(\mathbb{C} \ell(2 n+1)\) - Pauli spinors (\(\Psi^{\{3\}}\)) for \(\Gamma_{A}^{\{3\}}\).
\end{itemize}
If we consider a particular subalgebra of a higher-dimensional Clifford algebra, these components correspond to the spinors in the subalgebra space. And under the choice of different subalgebras, the \(2^{n}\)-component spinors \(\Psi_{1}^{(\iota)}\) and \(\Psi_{2}^{(\iota)}\) will describe different fermions, which implies it seems that different choices of subalgebra in the reducing process will lead to different effective theories.

We have reason to believe that these seemingly different effective theories should describe the same physics by a representation transformation with the mixing of the spinor bases. Because they are consistent in a higher-dimensional fundamental theory. These isomorphisms may be formally represented through a similarity transformation in the spinor space .

\section{Gamma matrices and spinors under $SO(n, 1)$}
Under a Lorentz transformation \(x \mapsto x^{\prime}\), the Dirac spinor transforms as
\begin{equation}\label{Lorentz_transformation}
\Psi^{\prime}\left(x^{\prime}\right)=S \Psi(x).
\end{equation}
An explicit expression for \(S\) is given by
\begin{equation}
S(\Lambda)=e^{-(i / 4) \Omega_{AB} \Sigma^{AB}},
\end{equation}
where \(\Omega_{AB}\) parameterize the Lorentz transformation, and \(\Sigma^{AB}\) are the \(4 \times 4\) matrices
\begin{equation} \label{Lorentz_transformation_op}
\Sigma^{AB}=\frac{i}{2}\left[\Gamma^{A}, \Gamma^{B}\right].
\end{equation}
These matrices can be interpreted as the intrinsic angular momentum of the Dirac field.

If two distinct sets of Gamma matrices that satisfy the Clifford relation are given, they can be related through a similarity transformation as
\begin{equation}
\Gamma'^{A} = S^{-1} \Gamma^{A} S.
\end{equation}
For the operator \(\Gamma^{M} D_{M}\) to remain invariant under a Lorentz transformation, the Gamma matrices need to undergo a transformation that corresponds to a contravariant vector with respect to their spacetime index, given by
\begin{equation}
S^{-1} \Gamma^{A} S = \Lambda^{A}_{~B} \Gamma^{B}.
\end{equation}
From Eq.~\eqref{Lorentz_transformation_op}, considering the matrix \(\Gamma^0\) is anti-hermitian and the other \(\Gamma\) matrices are hermitian, it follows that
\begin{equation}
\left(\Sigma^{AB}\right)^{\dagger}= \Gamma^0 \Sigma^{AB} \Gamma^0.
\end{equation}
We define the quantity \(\bar {\Psi}\) as
\begin{equation}
\bar {\Psi}={\Psi}^{\dagger} \Gamma_0.
\end{equation}
The transformed \(\bar{\Psi}^{\prime}\) can be expressed as
\begin{equation}
\bar{\Psi}^{\prime}(x^{\prime})=\Psi(x)^{\dagger} S(\Lambda)^{\dagger} \Gamma^{0}=\bar{\Psi}(x) e^{+(i / 4) \Omega_{AB} \Sigma^{AB}} =  \bar{\Psi}(x) S^{-1}.
\end{equation}
Thus,
\begin{equation}
\bar{\Psi}^{\prime}(x^{\prime}) \Psi^{\prime}(x^{\prime})= \bar{\Psi}(x) \Psi(x).
\end{equation}
In relativistic physics, it is \(\bar{\Psi} \Psi\), but not \(\Psi^{\dagger} \Psi\), that transforms as a Lorentz scalar. \(\bar{\Psi}\) represents the antiparticle of \(\Psi\). When mass is present, the left- and right-handed fields mix due to the equations of motion
\begin{equation}\label{Dirac Lorentz scalar}
\left(\Gamma^{M} D_{M}-m\right) \Psi^{\prime}(x^{\prime}, t^{\prime})=0,
\end{equation}
where
\begin{equation}
\Psi^{\prime}=S \Psi \quad \text{and} \quad \Gamma^{M}\equiv\Gamma^{A} E_A^M.
\end{equation}

%\bibliographystyle{JHEP}
%\bibliography{bibliography}

\acknowledgments
We are grateful to Zheng-Quan Cui and Yu-Peng Zhang for their valuable contributions to our discussions. This research was financially supported by the National Key Research and Development Program of China (Grant No.~2020YFC2201503), the National Natural Science Foundation of China (Grants No.~11875151 and No.~12247101), the 111 Project (Grant No.~B20063), and a research funding subsidy from Lanzhou City to Lanzhou University.

\providecommand{\href}[2]{#2}\begingroup\raggedright\endgroup

\end{document}